\documentclass[prc,aps,twocolumn,showpacs,letterpaper]{revtex4}
\begin{document}

\title{Fragment Production  in Non-central Collisions of 
Intermediate Energy Heavy Ions}

\author{B. Davin}
\author{R. Alfaro}
\author{H. Xu}
\author{L. Beaulieu}\altaffiliation{Present address: Universite Laval, Quebec, Canada.}
\author{Y. Larochelle}\altaffiliation{Present address: Universite Laval, Quebec, Canada.}
\author{T. Lefort}\altaffiliation{Present address: Universite de Caen, Caen, France.}
\author{R. Yanez}\altaffiliation{Present address: Universidad de Chile, Santiago, Chile.}
\author{S. Hudan}
\author{A.L. Caraley}
\author{R.T. de Souza}
\affiliation{
Department of Chemistry and Indiana University Cyclotron Facility, \\ 
Indiana University, Bloomington, IN 47405} 

\author{T.X. Liu}
\author{X.D. Liu}
\author{W.G. Lynch}
\author{R. Shomin}
\author{W.P. Tan}
\author{M.B. Tsang}
\author{A. Vander Molen}
\author{A. Wagner}\altaffiliation{Present address: Institute of Nuclear and Hadron Physics,
Forschungszentrum, Rossendorf, Dresden, Germany. }
\author{H.F. Xi}
\author{C.K. Gelbke}
\affiliation{
National Superconducting Cyclotron Laboratory and Department of
Physics and Astronomy,
Michigan State University, East Lansing, MI 48824}

\author{R.J. Charity}
\author{L.G. Sobotka}
\affiliation{
Department of Chemistry, Washington University, St. Louis, MO 63130}

\date{\today}

\begin{abstract}
The defining characteristics of fragment emission 
resulting 
from the non-central collision of $^{114}$Cd ions with $^{92}$Mo target 
nuclei at E/A = 50 MeV are presented. 
Charge correlations 
and average relative velocities for mid-velocity fragment 
emission exhibit significant differences when
compared to standard statistical decay. 
These differences associated with similar velocity dissipation 
are indicative of the influence of the
entrance channel dynamics on the fragment production process.

\end{abstract}

\maketitle

Peripheral and mid-central collisions of two heavy-ions at 
intermediate energies 
(20 MeV$\leq$E/A$\leq$50 MeV) result in the copious production of 
intermediate mass fragments (IMF: 3$\le$Z$\le$20). 
Whether these fragments originate from the surface instability of
non-spherical, transient geometries produced in 
nucleus-nucleus collisions \cite{neck1,Montoya94} 
or arise from proximity-enhanced, statistical decay of the projectile-like
and target-like fragments \cite{Botvina99} remains an open question. 
Fragment production as a result of dynamical surface 
instabilities would provide insight into the fragmentation 
mechanism of deformed nuclear 
matter at modest excitation energies. 
Alternatively, observation of the statistical 
decay of nuclear matter in the presence of an external long-range 
force represents a new and unexplored domain.  

At low incident energies, E/A$\le$10 MeV, nucleon transport between 
two colliding nuclei is Pauli-blocked,  allowing the projectile and target 
nuclei to remain approximately intact. 
Nucleons are exchanged {$\it via$} the ``neck'' (overlap region) 
connecting the two nuclei by a largely stochastic process \cite{Schroder80}.
Occasionally, dynamical neck rupture driven by surface instabilities can lead 
to production of fragments between the two reaction partners \cite{neck1}.
In contrast, at high incident energies, E/A$\ge$100 MeV, 
where nucleon-nucleon collisions
dominate, the ``participant'' region of overlap between the two nuclei 
becomes highly excited and rapidly disintegrates into a broad spectrum 
of nucleons and light nuclei \cite{Bowman73,Gosset77}. 
In this work, we are concerned with the origin of fragments at 
intermediate energies, E/A = 20-50 MeV, where experiments have observed
copious IMF yields at velocities  
between the projectile-like and target-like residues
\cite{Montoya94,Toke95,Lukasik97,Plagnol99,Larochelle97,Piantelli02}.
Since emission of such fragments is a robust prediction of transport 
theories \cite{Colonna93}, these ``neck fragments'' are often regarded as
dynamical in origin \cite{Milazzo01}. Recent theoretical and experimental analyses 
\cite{Botvina99,Mastinu96} however, have considered the statistical 
origin of such fragments. Thus, it is important to determine 
the defining characteristics associated with these fragments in order 
to elucidate the relevant mechanism. 

To explore the issues outlined above, we have 
examined peripheral and mid-central collisions in which
interaction of projectile and target nuclei  
results in the production of an excited 
projectile-like fragment (PLF$^*$) that can decay either
by evaporation or fission. If the creation of 
the PLF$^*$ and its subsequent decay 
are sufficiently decoupled, one expects the
decay of the PLF$^*$ to be isotropic in its own rest frame.
Alternatively, coupling between the formation of the PLF$^*$ and
its decay could result in an anisotropic decay pattern. 
Additionally, even if the formation 
and decay of the PLF$^*$ are decoupled, 
dynamical breakup, ({\it e.g.} neck rupture 
\cite{neck1,neck2}) 
or the proximity influence of the target-like fragment (TLF) 
may result in different
characteristics ({\it e.g.} yield, relative velocities) of fragments 
emitted forward and backward of the PLF$^*$.

The experiment was performed at the National Superconducting Cyclotron 
Laboratory at Michigan State University where a beam of $^{114}$Cd ions,  
accelerated by the K1200 cyclotron to E/A = 50 MeV, bombarded 
a 5.45 mg/cm$^2$ thick $^{92}$Mo target foil. 
Charged reaction products were detected by a setup 
which subtended 80\% of 4$\pi$. In the range 
2.1$^{\circ}$$\le\theta_{lab}\le$4.2$^{\circ}$, 
forward moving charged products were detected by a 300 $\mu$m thick,
segmented silicon ring counter (RC). The RC had 
4 separate quadrants of 16 independent annular sections on its front 
face and 16 pie-shaped sectors
on its back face and was thus capable of identifying
multiple charged particles. Behind each pie-sector, and covering the 
same solid angle, was 
a 2-cm thick CsI(Tl) crystal read-out with a photodiode. Identification
thresholds in the RC detector (RD) ranged from $\approx$9 MeV/A for 
Be to $\approx$25 MeV/A for Cd, 
with better than unit Z resolution for Z$\leq$48.
Fragment velocities were determined by 
assigning the A from the measured Z using the systematics of Ref. 
\cite{Summerer}. 
Reaction products emitted at 
larger angles (7$^{\circ}\le\theta_{lab}\le$168$^{\circ}$)
were detected by the silicon strip array LASSA \cite{LASSA}, and the Miniball-Miniwall 4$\pi$ array \cite{Miniball}.

\begin{figure} \vspace*{3.5in}
\includegraphics{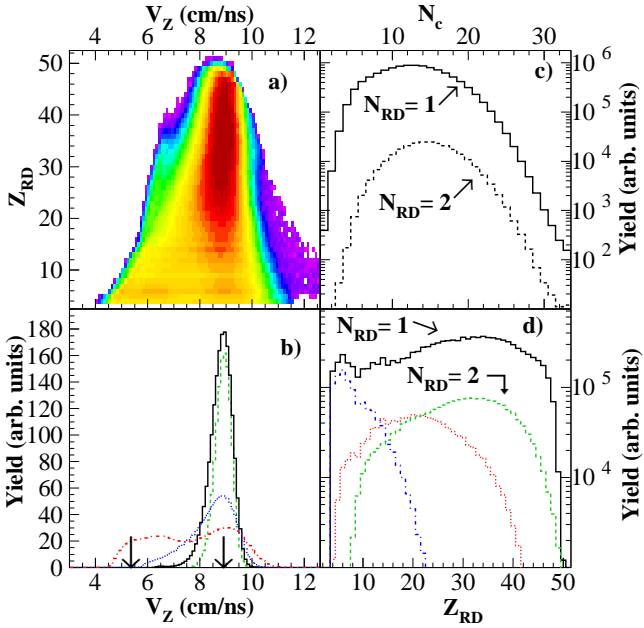}
\caption[]
{ 
a) Z$_{RD}$ {\it vs.} velocity for fragments detected in the ring 
detector (RD) associated with N$_c$ $\ge$ 3.
b) Fragment v$_{Z}$ distributions in the RD for 
different intervals in Z 
(40$\le$Z$\le$45 (dashed, green),
30$\le$Z$\le$35 (solid, black), 15$\le$Z$\le$20 (dotted, blue), 
or 4$\le$Z$\le$10 (dot-dash, red)). Arrows depict the
center-of-mass velocity (left) and beam velocity (right).
c) Multiplicity distributions associated with N$_{RD}$ $=$ 1 (solid) 
and  N$_{RD}$ $=$ 2 (dashed).
d) Z distribution of fragments measured in the RD when 
N$_{RD}$ $=$ 1 (solid, black) and  N$_{RD}$ $=$ 2 (dashed, green) 
and for the two
fragments individually (dots, red, larger fragment; dot-dash, blue, smaller 
fragment).} 
\end{figure}

Some general features of the reaction are depicted in Fig.~1.
Events were selected by the 
requirement that at least three charged particles 
were detected in the Miniball/Miniwall array and 
at least one charged particle
was detected in the RD.  
The relationship between the charge and  
velocity of the detected fragments in the RD is depicted in Fig.~1a. 
One observes a vertical ridge 
of fragments centered at approximately 9.0 cm/ns that increases
in width as Z decreases. 
In Fig.~1b, velocity 
distributions are displayed for fragments detected in the RD and selected
on different intervals in Z. 
When the
fragment detected in the RD is large, 40$\le$Z$\le$45 (dashed, green),   
30$\le$Z$\le$35 (solid, black), or 15$\le$Z$\le$20 (dotted, blue), 
the velocity distribution is peaked at 
$\approx$8.95 cm/ns.
This constancy in the velocity of the PLF has been attributed to decoupling 
of the overlap region from the non-overlap region (participant-spectator 
scenario) \cite{Westfall}. 
However, for the smallest 
fragments detected in the RD, 4$\le$Z$\le$10 (dot-dash, red), 
the distribution 
of velocities is bimodal, with peaks at velocities of $\approx$6.4 and 
$\approx$9.1 cm/ns.

\begin{figure} \vspace*{3.3in}
\includegraphics{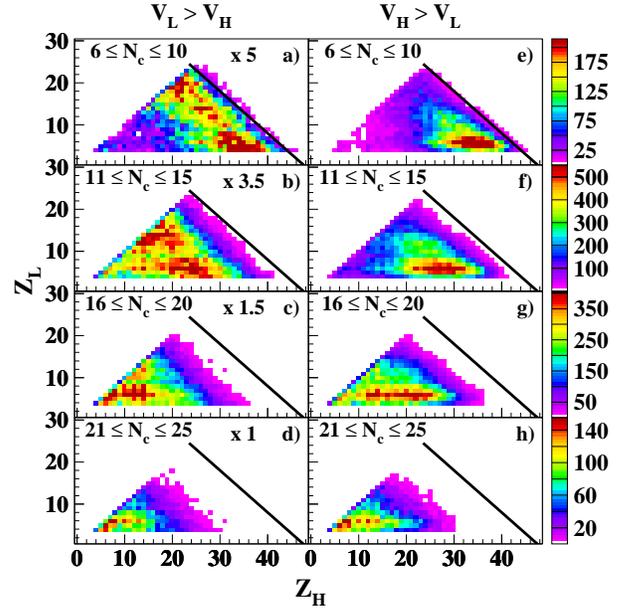}
\caption[]
{Charge correlation between the two fragments detected in the RD 
selected on both the magnitudes of the fragment velocities and N$_c$.} 
\end{figure}

	In order to study the decay properties of the PLF$^*$
produced in the peripheral and mid-central collisions we examine 
events in which two fragments (Z$\ge$4) are detected in the RD.
The high angular granularity of the RD allows us to selectively study the
binary decay of the PLF$^*$ along its flight direction. This 
selection has been shown to be a sensitive probe of the decay 
characteristics of the PLF$^*$ \cite{Bocage00}.
To begin we compare the charged particle multiplicity, N$_c$, distributions 
associated with detection of one (N$_{RD}$=1) or two (N$_{RD}$=2) fragments
in the RD as shown in Fig.~1c. When N$_{RD}$=1, 
$\langle$N$_c$$\rangle$=12.5 with a FWHM of 5.0. 
The multiplicity distribution associated with N$_{RD}$=2, 
does not differ substantially 
and has a mean value and FWHM of 14.8 and 4.0 respectively. 
On the basis of N$_c$ alone, these two cases appear similar.

The Z distributions of fragments in the ring detector, Z$_{RD}$, 
corresponding to  
N$_{RD}$=1 and N$_{RD}$=2, are shown in Fig.~1d. 
When N$_{RD}$=1, the Z distribution is broad and peaked 
at Z$_{RD}$$\approx$32. 
When N$_{RD}$=2, the distribution of total charge in the RD 
(dashed histogram, green) 
is similar to N$_{RD}$=1 (solid, black). The  individual Z distributions 
for the
larger fragment (dot, red) and smaller fragment (dot-dash, blue) are displayed 
for reference. The similarity of the total charge distribution when
N$_{RD}$=2, to the Z distribution for N$_{RD}$=1 suggests that these
events do not differ significantly from the N$_{RD}$=1 case. 

We have further separated the events
with N$_{RD}$=2
into two groups based upon the magnitudes of the fragment velocities.
In the first group, the smaller fragment is faster (v$_L$$>$v$_H$), while
in the second group the larger fragment is faster (v$_H$$>$v$_L$). 
If the decay of the PLF$^*$ is decoupled from its formation, 
and the decay is statistical,
the characteristics of these two groups of events should be identical.

The charge correlation between the two fragments detected in the 
RD is depicted in Fig.~2 as a function of N$_c$. 
In the left-hand column, the charge 
correlation between the two fragments for 
events in which v$_L$$>$v$_H$ is shown.
For the lowest multiplicity interval (Fig.~2a) a ridge of 
anti-correlated 
yield is observed -- consistent with the binary decay of a single 
parent fragment. 
For reference, the solid line 
corresponding to Z$_H$ + Z$_L$ = 48 is displayed. The shift of the 
ridge from 
this line indicates that the Z of the parent fragment, PLF$^*$, 
is on average slightly 
smaller (Z$_H$ + Z$_L$ $\approx$ 41) than the projectile atomic number. 
For the next multiplicity interval 
(Fig.~2b) a similar pattern is 
observed. The ridge of 
maximum yield has shifted to a smaller value (Z$_H$ + Z$_L$ $\approx$ 32).
This result is consistent with 
a less peripheral collision in an ablation/abrasion scenario or 
multi-particle emission from the PLF$^*$ before it splits into two 
fragments. 
We hypothesize that the observed 
binary pattern is fission-like in nature based upon lower energy
investigation of angular correlations of similar systems
\cite{Casini93,Stefanini95}.
For higher multiplicities, the anti-correlation pattern 
is no longer clearly evident. 
The pattern for events in which 
v$_H$$>$v$_L$, is shown in the right-hand column
of Fig.~2.
For all the multiplicity intervals displayed (panels e - h), super-imposed 
on a broad anti-correlated band, a horizontal ridge is discernible. 

The charge correlation patterns shown in Fig.~2 suggest a  
different behavior when v$_H$$>$v$_L$ as compared to v$_L$$>$v$_H$,
the first evidence we are selecting two different mechanisms. 
When v$_H$$>$v$_L$, the horizontal ridge indicates that 
the size of the smaller fragment is largely  
independent of the size of the larger fragment and has a most probable 
value of Z$_L$ $\approx$ 6. 
We conclude this insensitivity of the magnitude Z$_L$ to 
the magnitude of Z$_H$ is a defining 
feature of this process. 
Such behavior is consistent with the smaller 
fragment arising from the dynamical breakup of a neck-like structure 
where the most probable size of the smaller fragment is determined by the 
diameter of a cylindrical neck \cite{Montoya94}, 
or from sufficiently different statistical emission conditions as compared 
to the case when v$_L$$>$v$_H$. 
It is interesting to note that for the multiplicity interval 
6$\le$N$_c$$\le$10,
$\langle$Z$_L$$\rangle$ + 
$\langle$Z$_H$$\rangle$ is essentially the same 37.6 and 38.3 for 
v$_L$$>$v$_H$ and v$_H$$>$v$_L$ respectively. This result is consistent 
with the N$_c$ selection determining the impact parameter and hence the 
size of the fragmenting system in both cases.
The difference between these 
{\it charge correlation patterns} observed for the two cases represents a 
distinguishing feature of the mid-velocity fragment emission not realized in 
earlier work \cite{Montoya94,Bocage00}.

\begin{figure} \vspace*{3.5in}
\includegraphics{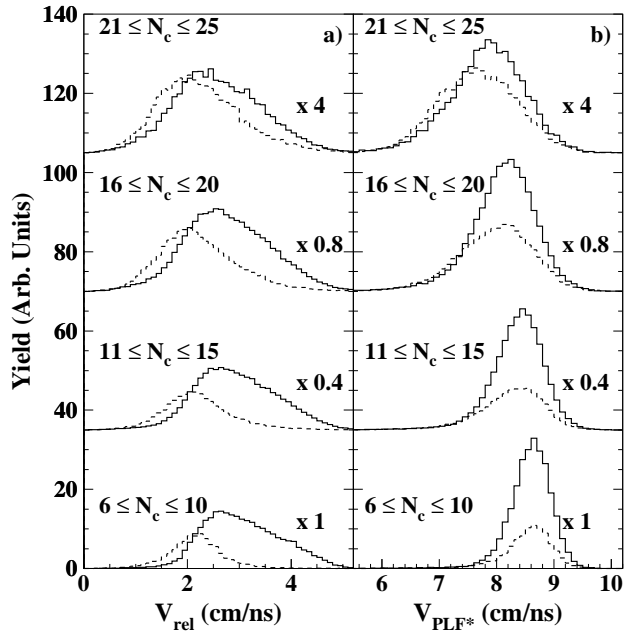}
\caption[]
{a) Distributions of relative velocity between the two RD fragments for 
v$_H$$>$v$_L$ (solid histogram) and v$_L$$>$v$_H$ (dashed histogram) for 
the indicated selections on the charged particle multiplicity, N$_C$.
b) Distributions of v$_{PLF^*}$, the center-of-mass velocity of the two
fragments detected in the RD. Solid histograms represent the case for 
v$_H$$>$v$_L$ and dashed histogram the case v$_L$$>$v$_H$. 
} 
\end{figure}

We have further examined the differences between the  
v$_H$$>$v$_L$ and v$_L$$>$v$_H$ cases by comparing the distributions of the
relative velocity between these two fragments, $\vec{v}_{rel}$ 
= $\vec{v}_H - \vec{v}_L$
and the distributions
of their center-of-mass velocity, v$_{PLF^*}$, for selected cuts on the 
charged particle multiplicty, N$_c$. Evident in Fig.~3a is the fact that the
v$_{rel}$ distribution is broader with a higher mean value when
v$_H$$>$v$_L$ as compared to v$_L$$>$v$_H$. This difference is largest 
when 6$\le$N$_c$$\le$10. In contrast to the differences observed in v$_{rel}$,
the distributions in v$_{PLF^*}$ (Fig.~3b) for both cases are remarkably 
similar in both their average value and their width. Two trends are evident
in Fig.~3b. With increasing N$_c$, the mean value of 
v$_{PLF^*}$ decreases, and the width of the
v$_{PLF^*}$ distribution also increases systematically. 
If one relates the v$_{PLF^*}$ to the velocity damping of the PLF and 
consequently its excitation, the trend of a decrease in the mean value of 
v$_{PLF^*}$ with increasing N$_c$ is reasonable. 

\begin{figure} \vspace*{3.5in}
\includegraphics{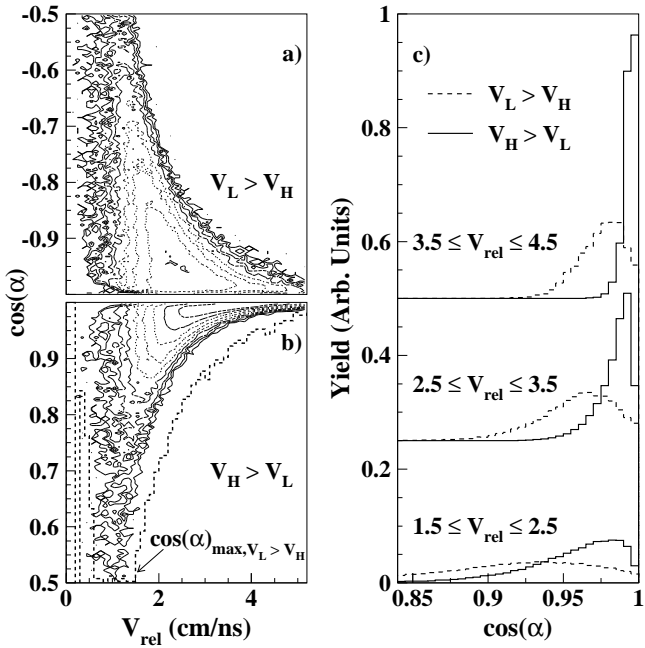}
\caption[]
{Angular correlation between v$_{rel}$ and v$_{PLF^*}$ for v$_L$$>$v$_H$ 
(panel a) and v$_H$$>$v$_L$ (panel b). The dashed line in panel b) depicts 
the detector acceptance for the v$_L$$>$v$_H$ case. Angular distributions for
v$_H$$>$v$_L$ (solid histogram) and v$_L$$>$v$_H$ (dashed histogram)
for different intervals in v$_{rel}$ are shown in panel c).

} 
\end{figure}

The angular distribution between v$_{rel}$ and v$_{PLF^*}$ is examined in 
Fig.~4 for 6$\le$N$_c$$\le$25 as a function of v$_{rel}$. 
We define cos($\alpha$) = 
($\vec{v}_{PLF^*}).(\vec{v}_{rel}$)/(v$_{PLF^*}$v$_{rel}$)). These two 
distributions are quantitatively different. When v$_H$$>$v$_L$ (Fig.~4b),  the 
cos($\alpha$) distribution is narrower than when 
v$_L$$>$v$_H$ (Fig.~4a). 
These two-dimensional distributions   
reflect both the true angular distribution and the influence of the 
detector acceptance. Depicted as a 
dashed line in panel b is the maximum value of cos($\alpha$) for 
v$_L$$>$v$_H$ which we designate cos($\alpha$)$_{max,v_L > v_H}$,
as a function of v$_{rel}$. The fact that the entire measured 
distribution for v$_H$$>$v$_L$ is less than
cos($\alpha$)$_{max,v_L > v_H}$, 
suggests that the limited angular coverage
of the RD {\it does not} present a significant problem for
the v$_H$$>$v$_L$ case. Since the geometric acceptance depends on v$_{PLF^*}$,
we have examined these same distributions for more restrictive conditions by
selecting on 6$\le$N$_c$$\le$10 so as to select a more defined v$_{PLF^*}$.
The result of this more restrictive analysis supports the conclusion that
for v$_H$$>$v$_L$ the experimental angular coverage is sufficient.

In order to more
quantitatively compare the angular distributions between the two velocity 
cases, we show in Fig.~4c distributions in cos($\alpha$) for different 
intervals in v$_{rel}$. Solid histograms represent the case
v$_H$$>$v$_L$ and dashed histograms the case v$_L$$>$v$_H$. For all three 
intervals of v$_{rel}$ shown, the distribution for v$_H$$>$v$_L$ 
is more sharply peaked towards 
cos($\alpha$)=1 than the v$_L$$>$v$_H$ distribution,
indicating alignment of the two fragments is more probable when 
v$_H$$>$v$_L$. This 
observed angular correlation is 
consistent with the work of \cite{Bocage00}.

Based upon the charge correlations observed in Fig.~2, we have 
further subdivided both types of events 
according to the magnitude of Z$_L$: Z$_L$$<$9
and Z$_L$$>$9. The kinetic characteristics of the fragment pairs selected in this way are shown in 
Fig.~5. 
The average velocity  of the PLF$^*$, $\langle$v$_{PLF*}$$\rangle$, 
is presented for the different cases in Fig.~5a. 
One observes that all four cases are clustered in a band, consistent with the
behavior in Fig.~3b, with a 
lower velocity than the N$_{RD}$=1 case. 
The similar magnitude of $\langle$v$_{PLF*}$$\rangle$ suggests 
that the velocity damping (excitation) of the system in all four cases 
is similar and larger than the velocity damping (excitation) in the 
N$_{RD}$=1 case. This conclusion
is further borne out by examining the multiplicity of light charged particles
detected. For the most peripheral
collisions studied, {\it i.e.}  the multiplicity interval 
6$\le$N$_c$$\le$10, $\langle$N$_H$$\rangle$ and $\langle$N$_{He}$$\rangle$ 
are 3.6 and 2.2 when v$_H$$>$v$_L$ and 3.6 and 2.3 when v$_L$$>$v$_H$.
The similarity of the magnitude of v$_{PLF*}$, for both velocity cases, as 
well as the similar light charged particle multiplicities 
suggests that the excitation
(and spin) of the PLF$^*$ is essentially the same in both velocity cases.

\begin{figure} \vspace*{3.4in}
\includegraphics{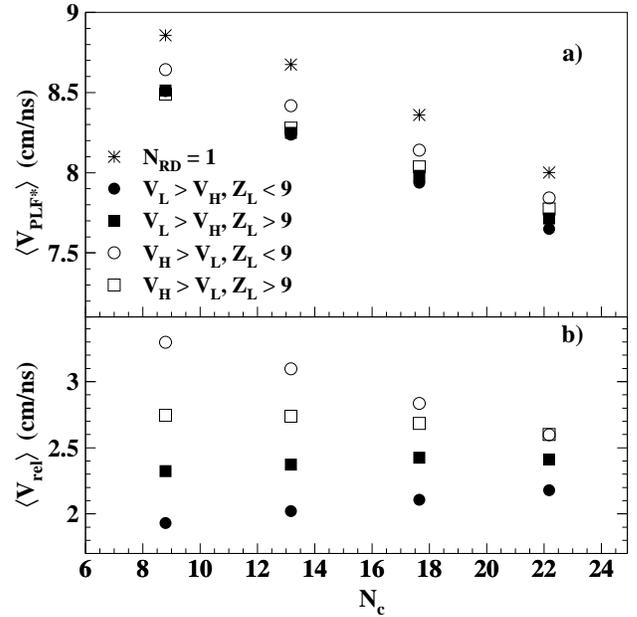}
\caption[]
{
a) Dependence of center-of-mass velocity of the two RD
fragments on multiplicity. 
b) Relation between the  relative velocity of the two fragments in the 
RD and N$_c$.
} 
\end{figure}

We have also examined 
the dependence of the average relative velocity of the two fragments, 
$\langle$v$_{rel}$$\rangle$, on N$_c$ 
as shown in Fig.~5b. 
For v$_L$$>$v$_H$ and Z$_L$$>$9, (filled squares), 
$\langle$v$_{rel}$$\rangle$ 
is relatively
constant as a function of N$_c$, suggesting that the 
decay of the PLF$^*$ is sufficiently decoupled from its formation.
The average relative velocity increases slightly with N$_c$ when 
v$_L$$>$v$_H$ and Z$_L$$<$9, (filled circles).
More symmetric splits (Z$_L$$>$9) 
exhibit a somewhat larger 
$\langle$v$_{rel}$$\rangle$ ($\approx$2.3-2.4 cm/ns) than 
more asymmetric (Z$_L$$<$9) 
splits ($\approx$1.9-2.2 cm/ns). 
Both these values of 
$\langle$v$_{rel}$$\rangle$ 
and the N$_c$ independence are consistent with the 
fission of a PLF$^*$ (Viola fission TKE 
systematics \cite{Viola85}) following an initial interaction which determines
the N$_c$. 
When v$_H$$>$v$_L$ and Z$_L$$>$9  (open squares) a similar independence of 
$\langle$v$_{rel}$$\rangle$ on 
N$_c$ is 
observed although the magnitude of 
$\langle$$v_{rel}$$\rangle$ is slightly larger ($\approx$2.6-2.8 cm/ns). 
This constancy of 
$\langle$v$_{rel}$$\rangle$  
suggests that these events also arise from a 
mechanism similar to the two-stage fission-like behavior observed for 
v$_L$$>$v$_H$.

In marked contrast to 
the previously described fission-like  behavior 
is the behavior for Z$_L$$<$9 
when v$_H$$>$v$_L$  (open circles). 
At low N$_c$ for this case, 
$\langle$v$_{rel}$$\rangle$  
is considerably higher than in the fission-like case. 
With increasing N$_c$ (centrality) 
the $\langle$v$_{rel}$$\rangle$ 
for this case 
decreases. 
The monotonic decrease of 
$\langle$v$_{rel}$$\rangle$, 
in contrast to the
near constant behavior observed for the other cases, suggests that 
in this case the fragment
emission arises from a different mechanism -- one coupled to
formation of the PLF$^*$. 
Large relative velocities have 
recently been reported \cite{Bocage00} over a broad range of charge asymmetry
for this additional component; however, the present work focuses on the most
extreme asymmetries for which we determine the effect is the largest. 
Moreover, we determine, for the first time, that these large differences in
$\langle$v$_{rel}$$\rangle$ are associated with essentially the same 
dissipation.

As the largest variations in 
$\langle$v$_{rel}$$\rangle$ are evident for 
6$\le$N$_c$$\le$10, we now focus on this multiplicity interval.
Comparison of $\langle$v$_{rel}$$\rangle$ in Fig.~5 between
Z$_L$$<$9 and Z$_L$$>$9  for either v$_L$$>$v$_H$ or v$_H$$>$v$_L$  
suggests a dependence of $\langle$v$_{rel}$$\rangle$ on Z$_L$.
In Fig.~6a, one observes that the two velocity cases manifest 
different trends of  
$\langle$v$_{rel}$$\rangle$ 
with increasing Z$_L$. 
For events selected with v$_L$$>$v$_H$ (closed circles), 
$\langle$v$_{rel}$$\rangle$ 
is approximately constant for Z$_L$$\ge$7, consistent with 
the Viola fission systematics (solid line) \cite{Viola85}, 
{\it i.e.} a Coulomb dominated separation. 
For Z$_L$$\le$6, the values of  
$\langle$v$_{rel}$$\rangle$ 
are somewhat lower than expected 
possibly due to reduced efficiency for triggering on fast, light
fragments in the RD. In addition, the influence of secondary decay effects,
increases in magnitude with decreasing Z$_L$.  
The error bars shown reflect the uncertainty due to the 
deduced mass of the fragment.

\begin{figure} \vspace*{3.3in}
\includegraphics{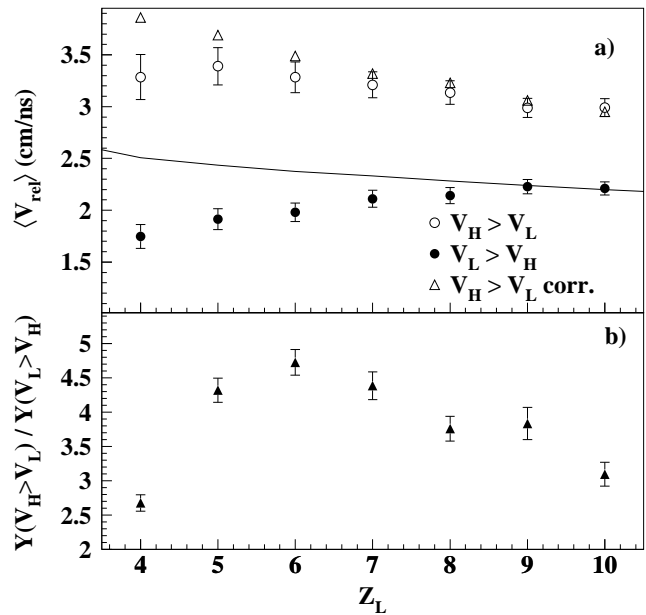}
\caption[]
{
a) Dependence of the relative velocity of the two fragments 
detected in the 
RD on the charge of the smaller fragment, Z$_L$. 
b) Ratio of the yield of Z$_L$ for the two cases v$_H$$>$v$_L$ and 
v$_L$$>$v$_H$, when 6$\le$N$_c$$\le$10.
} 
\end{figure}

For the case of v$_H$$>$v$_L$ (open circles), 
one observes a significantly larger magnitude for 
$\langle$v$_{rel}$$\rangle$, 
which decreases as Z$_L$ increases. The 
enhancement of relative velocity above the Coulomb lower limit 
(solid line) is approximately 40$\%$ for Z=4. 
We conclude this enhancement is another defining characteristic of
mid-rapidity fragment emission. 
Correction for the finite acceptance and thresholds of the RD result
in the open triangles displayed in Fig. 6. These restrictions do not change the
qualitative description of the data and only result in enhancing the observed 
trends.

Insight into the mechanism of mid-rapidity fragment production 
is provided by considering the possible sources of the 
the observed enhancement in 
$\langle$v$_{rel}$$\rangle$. Fragments acquire their velocities
from any of three possible sources: 
Coulomb repulsion, thermal energy, or collective
motion. The difference 
in $\langle$v$_{rel}$$\rangle$ observed in Fig.~6a 
cannot be attributed to Coulomb repulsion, nor to significant thermal 
differences due to the N$_c$ selection (and supported by the average 
multiplicity of light charged particles). Thus, the significant 
enhancement of 
$\langle$v$_{rel}$$\rangle$ 
is most likely collective in origin and due to 
coupling to the 
considerable collective motion present in the entrance channel. 

In Fig.~6b we present the ratio of 
the measured yields of v$_H$$>$v$_L$ to v$_L$$>$v$_H$ 
as a function of Z$_L$ (closed triangles).
Over the entire range of Z$_L$ displayed, 
the yield for the v$_H$$>$v$_L$ case is larger than the v$_L$$>$v$_H$ case 
by a factor of 2.6-4.7 and reaches a maximum at Z$_L$=6. 
This enhancement of yield at velocities intermediate between the PLF and TLF 
is comparable to the enhancement observed 
previously \cite{Montoya94}.

\begin{figure} \vspace*{3.3in}
\includegraphics{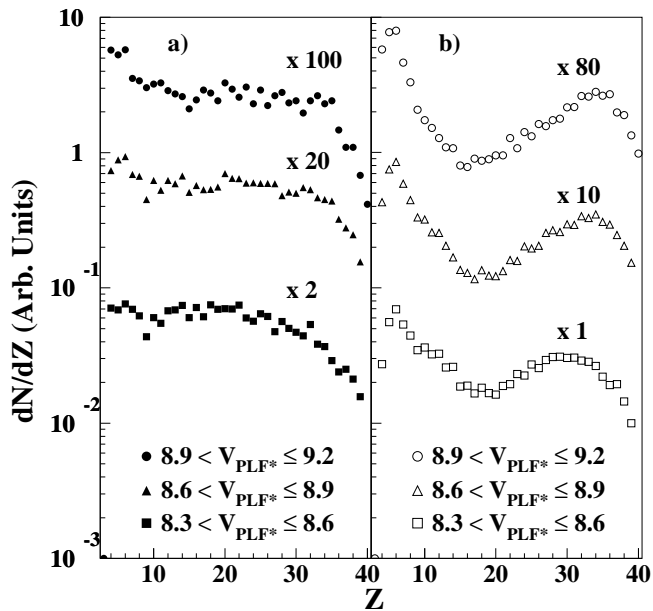}
\caption[]
{Z distributions when
6$\le$N$_c$$\le$10
for 8.9$<$v$_{PLF^*}$(cm/ns)$\le$9.2,  
8.6$<$v$_{PLF^*}$(cm/ns)$\le$8.9,  and
8.3$<$v$_{PLF^*}$(cm/ns)$\le$8.6,  when 
v$_L$$>$v$_H$ (solid symbols) and 
v$_H$$>$v$_L$ (open symbols).
} 
\end{figure}

In order to characterize the nature of the coupling to the entrance channel,
we examine the Z 
distribution for peripheral collisions (6$\le$N$_c$$\le$10) selected
on the  velocity of the PLF$^*$.
For this multiplicity interval the most probable value of 
Z$_H$+Z$_L$ is $\approx$41 in both cases.
The velocity distribution of the PLF$^*$ for both
v$_H$$>$v$_L$ and v$_L$$>$v$_H$ is 
approximately gaussian with 
$\langle$v$_{cm}$$\rangle$=8.6, 8.5 cm/ns and a FWHM=0.41, 0.50 cm/ns 
for v$_H$$>$v$_L$ and v$_L$$>$v$_H$, respectively. 
A decrease in v$_{PLF^*}$ is related to an increase in the velocity 
dissipation and consequently
an increase in the excitation of the PLF$^*$ and possibly 
a change in its spin.
In Fig. 7 we examine 
the Z distribution selected on progressively 
decreasing values of v$_{PLF^*}$
for both v$_L$$>$v$_H$ (solid symbols)
and v$_H$$>$v$_L$ (open symbols).
One observes that
for the fission-like case, {\it i.e.} v$_L$$>$v$_H$ (closed symbols),
for the case of the least dissipation 8.9$<$v$_{PLF*}$(cm/ns)$\le$9.2, the Z 
distribution is slightly asymmetric. With increasing dissipation 
the distribution becomes relatively flat. 
An asymmetric distribution can be understood in terms of the fissility of
a light nucleus and the Businaro-Gallone point \cite{Bocage00,Charity88}.
As the transition from an asymmetric to a symmetric Z distribution 
is more sensitive to the spin of the decaying nucleus as compared to its
excitation, it is likely that the transition 
observed for the fission-like case
is due to increased spin of the
PLF$^*$ as v$_{PLF*}$ decreases.  
For a light system, such as Z=40, with low spin
one expects an asymmetric Z distribution;
with increasing spin, approaching the critical angular momentum
allowed for a rotating nucleus,
a transition to an increasingly flat Z distribution is expected
\cite{Charity88}.
One can alternatively relate the asymmetric component of the Z distribution
to evaporation and the symmetric component to fission. 
While excitation alone favors evaporation, 
inclusion of spin enhances the importance of the fission channel.
The transition from a slightly asymmetric distribution to a flat one with 
increasing velocity damping can be related to the
relative importance of symmetric fission as compared to evaporation 
and consequently the excitation and spin of the decaying system. 
 
In contrast to the behavior
observed for the fission-like case, when 
v$_H$$>$v$_L$ (open symbols) the Z distribution is always asymmetric
and maintains its asymmetry over the same interval in v$_{PLF^*}$ during
which the v$_L$$>$v$_H$ makes the transition from slightly asymmetric
to flat.
This behavior of the Z distribution asymmetry when v$_H$$>$v$_L$ suggests that 
the breakup in this case is driven by factors other than excitation or spin.
The observation of dynamically driven fragment production in {\it hot} 
ternary fission \cite{Yanez99} may indicate a common origin of 
mid-velocity fragments.

In summary, examination of fragment emission forward and backward of an excited
projectile-like fragment  
formed in the peripheral or mid-central collision of two heavy-ions
reveals two types of decay processes.
One process is consistent with the standard statistical 
fission-like decay while the other, associated with mid-velocity
fragment emission, clearly reflects the influence of the entrance channel, 
hence is a dynamical process. 
These two processes exhibit {\it substantially different charge correlations}.
Focusing on the most peripheral collisions,
we have further compared these two decay modes on the
basis of the dissipated velocity and we find that {\it for essentially 
the same 
velocity dissipation the $\langle$v$_{rel}$$\rangle$ and
the Z distributions differ markedly}.
Neither excitation nor spin of the PLF$^*$ can explain either 
the asymmetry of the 
Z distribution for the dynamical component or the persistence of this 
asymmetry with increased velocity damping. 

	We would like to thank the staff at MSU-NSCL for
providing the high quality beams which made this experiment possible.
One of the authors (RdS) is grateful to 
Commissariat a l' Energie Atomique and G.A.N.I.L. (France) 
for support enabling this work during a sabbatical. 
This work was supported by the
U.S. Department of Energy under DE-FG02-92ER40714 (IU), 
DE-FG02-87ER-40316 (WU) and the
National Science Foundation under Grant No. PHY-95-28844 (MSU).\par


\end{document}